

\input phyzzx

\def\PLB{ \sl Phys. Lett. \bf B}
\def\NPA{ \sl Nucl. Phys. \bf A}
\def\NPB{ \sl Nucl. Phys. \bf B}

\def\PRD{ \sl Phys. Rev.  \bf D}
\def\PRL{ \sl Phys. Rev. Lett. \bf}

\def\MPL{ \sl Mod. Phys. Lett. \bf}
\def\ZPC{ \sl Zeit. Phys. \bf C}

\def\LNC{ \sl Lett. Nuovo Cim. \bf}
\def\d { {\rm d} }

\def\tr{ {\rm tr} }
\def\tanh{ {\rm tanh} }
\def\sech{ {\rm sech} }
\def\L { {\cal L} }
\def\ZZ { {\bf Z} }
\def\tshalf{{\textstyle{1\over 2}}}
\def\tsfourth{{\textstyle{1\over 4}}}
\def\half{{1\over 2}}

\def\eps{\epsilon}
\def\lam{\lambda}
\def\sig{\sigma}

\def\del{\partial}
\def\vphi{\varphi}
\def\zo{ {(0)} }
\def\two{$(1\!+\!1)$}
\def\four{$(3\!+\!1)$}
\def\pone{ p_1 }
\def\wkb{{\scriptscriptstyle \rm WKB}}
\def\eff{ {\rm eff} }
\def\Leff { \L_\eff  }
\def\Eeff { E_\eff  }
\def\Seff { S_\eff  }
\def\Ltwo{  \L_\eff^{(2)} }
\def\Q { Q}
\def\xcl{ x_{\rm cl} }
\def\xqu{ x_0 }
\def\zqu{ z_0 }
\def\ksig{ k_\sig }
\def\wsig{ w_\sig }
\def\Vsig{ V_\sig }
\def\deltasig{ \delta_\sig }
\def\epssig{ \eps_\sig }
\def\psisig{ \psi_\sig }
\def\epssiglam{ \eps_{\sig\lam} }
\def\Delt { \Delta }
\def\Deltt {  U }
\def\Gamtwo{ \Gamma^{(2)}_{\sigma \sigma} }
\def\psii{ \psi^{i}}
\def\psiibar{{\overline \psi}^{i}}

\def\psievensig{\psi_{\sig,{\rm even}}}
\def\psioddsig{\psi_{\sig,{\rm odd}}}
\def\deven { \delta_{\rm even} }
\def\dodd { \delta_{\rm odd}}
\def\devensig { \delta_{\sig,{\rm even}} }
\def\doddsig { \delta_{\sig,{\rm odd}} }
\def\dE { \delta E }
\def\dL { \delta \L }
\def\QWKB {\Q^\wkb}
\def\Ecl { E_{\rm cl} }
\def\Edis { E_{\rm disc} }
\def\Edisone { E_{\rm disc}^{(1)} }
\def\Edistwo { E_{\rm disc}^{(2)} }
\def\Econ  { E_{\rm cont} }
\def\bkphi{\,[\vphi]\, }
\def\bkpphi{\,[\phi]\, }
\def\bkphicl{\,[\vphi_{\rm cl}]\, }
\def\vphicl{ \vphi_{\rm cl} }
\def\cancel#1#2{\ooalign{$\hfil#1\mkern1mu/\hfil$\crcr$#1#2$}}
\def\slash#1{\mathpalette\cancel{#1}}
\def\pslash{ \slash{p}}
\def\delslash{ \slash{\partial} }
\def\Dslash{\slash{D}}

\def\logdet{  \log \det \left( i\delslash - g \vphi \right) }
\def\logdetv{  \log \det \left( i\delslash - g v \right) }
\def\logdetD{  \log \det \left( i\Dslash \right) }
\def\scale{ \left( N \over \xcl \right) }
\def\tscale{ (N / \xcl) }
\def\gsim{\mathrel{\raise.3ex\hbox{$>$\kern-.75em\lower1ex\hbox{$\sim$}}}}
\def\lsim{\mathrel{\raise.3ex\hbox{$<$\kern-.75em\lower1ex\hbox{$\sim$}}}}
%
\REF\solitons{
	   S.~Coleman,
            \it Aspects of Symmetry
            \rm (Cambridge Univ.~Press, Cambridge, 1985), ch.~6;\nextline
           R.~Rajaraman,
            \it Solitons and Instantons
            \rm (North-Holland, Amsterdam, 1982).}
\REF\sG{   S.~Coleman,
             \PRD 11 \rm (1975) 2088;\nextline
	   S.~Mandelstam,
             \PRD 11 \rm (1975) 3026.}
\REF\unstable{
	   P.~Vinciarelli,
             \LNC 4 \rm (1972) 905; \nextline
	   T.~D.~Lee and G.~Wick,
             \PRD 9 \rm (1974) 2291; \nextline
	   A.~Chodos, R.~Jaffe, K.~Johnson, C.~Thorn, V.~Weisskopf,
             \PRD  9 \rm (1974)~3471;\nextline
           M.~Creutz,
             \PRD 10 \rm (1974) 1749; \nextline
           M.~Creutz and K.~Soh,
             \PRD 12 \rm (1975) 443; \nextline
           W.~Bardeen, M.~Chanowitz, S.~Drell, M.~Weinstein, T.~Yan,
             \PRD 11~\rm(1975)~1094; \nextline
           R.~Friedberg and T.~D.~Lee,
             \PRD 15 \rm (1977) 1694;
             \PRD 16 \rm (1977) 1096;
             \PRD 18 \rm (1978) 2623.}
\REF\CL  { D.~Campbell and Y.-T.~Liao,
             \PRD 14 \rm (1976) 2093.}
\REF\MWZ{  R.~MacKenzie, F.~Wilczek, and~A. Zee,
             \PRL  53 \rm (1984) 2203; \nextline
           R.~MacKenzie,  Ph.~D.~thesis,
             ITP preprint NSF-ITP-84-135; \nextline
           R.~MacKenzie,
             \MPL A7 \rm (1992) 293.}
\REF\DF{   E.~D'Hoker and E.~Farhi,
	      \NPB 241 \rm (1984) 109;
	      \NPB 248 \rm (1984)~77. }
\REF\Davis{ R.~L.~Davis,
              \PRD 38 \rm (1988) 3722.}
\REF\Kaplan{ D.~Kaplan,
              \PLB 235 \rm (1990) 163.}
\REF\MP{   R.~MacKenzie and W.~Palmer,
              \PRD 42 \rm (1990) 701.}
\REF\BN {  J.~Bagger and S.~Naculich,
	     \PRL 67 \rm (1991) 2252;  \nextline
             \PRD 45 \rm (1992) 1395.}
\REF\BD{ T.~Banks and A.~Dabholkar,
             Rutgers preprint, RU-92-09.}
\REF\Peris{ S.~Peris,
	      \PLB 251 \rm (1990) 603.}
\REF\LargeN{A somewhat different large-$N$ limit is used in: \nextline
   	   M.~Einhorn and G.~Goldberg,
               \PRL 57 \rm (1986) 2115; \nextline
           K.~Aoki,
               \PRD 44 \rm (1991) 1547; \nextline
           K.~Aoki and S.~Peris,
	      preprint UCLA/92/TEP/32, OHSTPY-HEPT-92-008.}
\REF\gradient{
	   I.~Aitchison and C.~Fraser,
            \PLB 146 \rm (1984) 63;
            \PRD 31 \rm~(1985)~2605; \nextline
           C.~Fraser,
            \ZPC 28 \rm (1985) 101; \nextline
           L.-H.~Chan,
            \PRL 54 \rm (1985) 1222;
            \bf  56 \rm (1985) 404(E);
            \bf 55 \rm (1985) 21; \nextline
           O.~Cheyette,
            \PRL 55 \rm (1985) 2394; \nextline
           J.~Zuk,
            \ZPC 29 \rm (1985) 303.}
\REF\DHN{  R.~Dashen, B.~Hasslacher and A.~Neveu,
             \PRD 10 \rm (1974) 4114; \nextline
             \PRD 10 \rm (1974) 4130;
             \PRD 12 \rm (1975) 2443.}
\REF\KPR{  G.~Ripka and S.~Kahana,
             \PLB 155 \rm (1985) 327;\nextline
  	   S.~Kahana, R.~Perry, and G.~Ripka,
             \PLB 163 \rm (1985) 37;\nextline
           R.~Perry,
             \NPA 467 \rm (1987) 717.}
\REF\WK{   D.~Wasson and S.~Koonin,
             \PRD 43 \rm (1991) 3400.}
\REF\JR{   R.~Jackiw and C.~Rebbi,
              \PRD 13 \rm (1976) 3398.}
\REF\CY{   S.-J.~Chang and T.-M.~Yan,
             \PRD 12 \rm (1975) 3225.}
\REF\COLSYS{
	   U.~Ascher, J.~Christiansen, and R.~Russell,
             \sl ACM Trans. Math. Sftw.
             \bf 7 \rm (1981) 223.}
\REF\super{ P.~DiVecchia and S.~Ferrara,
             \NPB  130 \rm (1977) 93; \nextline
	    J.~Hruby,
             \NPB  131 \rm (1977) 275.}
\overfullrule=0pt
\nopubblock
\line{\hfil JHU-TIPAC-920014}
\line{\hfil July, 1992}
\titlepage
\title{ {\seventeenrm \bf Quantum Kinks: Solitons at Strong Coupling} }
\author{Stephen G. Naculich\footnote\ast{
        NACULICH @ CASA.PHA.JHU.EDU }}
\address{Department of Physics and Astronomy     \break
        The Johns Hopkins University             \break
        Baltimore, MD 21218}
\abstract{
We examine solitons in theories with heavy fermions.
These ``quantum'' solitons differ dramatically
from semi-classical (perturbative) solitons
because fermion loop effects are important
when the Yukawa coupling is strong.
We focus on kinks in a $(1+1)$--dimensional $\phi^4$ theory
coupled to fermions;
a large-$N$ expansion is employed
to treat the Yukawa coupling $g$ nonperturbatively.
A local expression for the fermion vacuum energy
is derived using the WKB approximation for the Dirac eigenvalues.
We find that fermion loop corrections increase the energy of the kink
and (for large $g$) decrease its size.
For large $g$,
the energy of the quantum kink is proportional to $g$,
and its size scales as $1/g$,
unlike the classical kink;
we argue that these features are generic to quantum solitons
in theories with strong Yukawa couplings.
We also discuss the possible instability of fermions
to solitons.
}
\endpage

\chapter{Introduction}

Topological solitons,
despite their inherently nonperturbative character,
are typically studied semi-classically,
that is,
in a perturbative expansion in the coupling constants [\solitons].
The first term in this expansion,
the classical soliton,
is the solution to a nonlinear classical field equation.
This solution is nonperturbative
because its energy diverges as the coupling constants---which
parametrize the nonlinearity---vanish.
Perturbative corrections to the soliton are important:
they split the degeneracies of the classical solution
resulting from Poincar\'e and internal symmetries,
and project the solitons onto
eigenstates of momentum, angular momentum, and charge.
If the coupling constants are small, however,
corrections to the shape and energy of the soliton are small,
and the classical description of the soliton is essentially accurate.

If the couplings are large, on the other hand,
there is no reason to expect
the quantum soliton states to resemble
the classical solitons, at least quantitatively.
In general, the strong coupling behavior of solitons
in a quantum field theory is not well known.
One notable exception is the sine-Gordon kink in
\two~dimensions;
because of the equivalence of the sine-Gordon theory to the
massive Thirring model [\sG],
the sine-Gordon kink at strong coupling
becomes a weakly-coupled fermion in the Thirring model,
which is well described by perturbation theory.

In this paper we study strongly-coupled solitons more generally,
when such a fortuitous equivalence does not arise.
We focus in particular on solitons
in theories with large Yukawa couplings.
One motivation for doing so is the following.
Fermions can acquire mass through a Yukawa
coupling to a scalar field
with nonvanishing vacuum expectation value.
Solitons in such theories often carry (possibly fractional) fermion number.
It has recurrently been suggested
that when the Yukawa coupling is large
such a soliton may have less energy
than a fermion in a constant scalar field background;
consequently, fermions may be unstable
to the formation of solitons [\unstable--\BD].
To determine whether this is so, however,
one must know the form and energy of solitons
in a strongly-coupled theory,
which may differ appreciably from classical solitons.
Indeed, we expect fermion loop corrections to
significantly affect the solitons
when the Yukawa coupling is large.

One means of studying a strongly-coupled Yukawa theory
is through a large-$N$ expansion [\MWZ,\BN--\LargeN].
To leading order in $1/N$, the theory can be solved
for arbitrary values of the Yukawa coupling.
This expansion captures some of
the strong-coupling behavior of the theory,
which one hopes is representative
even when $N$ is not large.
To carry out this expansion,
we introduce $N$ fermion flavors
and choose the $N$-dependence of the couplings
so that the theory has a sensible $N \to \infty$ limit,
with only fermion loops contributing to Green functions
to leading order in $1/N$.
The total contribution of the fermion loops
can be summed in closed form to give
the exact large-$N$ effective action
$$
\Seff \bkpphi = S \bkpphi - iN \logdetD,
\eqn\eelargeNeffact
$$
where $ S \bkpphi $ is the classical scalar field action
and $\Dslash$ is the Dirac operator in the presence of
the field $\phi$.

Solitons in this large-$N$ theory
are $c$-number configurations of the scalar fields;
scalar field fluctuations are suppressed
because scalar loops do not contribute to the effective
action to leading order in $1/N$.
The shape of the large-$N$ soliton
differs from the classical soliton, however,
since it extremizes not the classical action
but the effective action \eelargeNeffact.
The fermion loop contribution significantly alters
the form of the soliton when the Yukawa coupling is large.
In this regime,
where quantum effects are so important,
the large-$N$ soliton
is truly a ``quantum soliton.''

To determine the form of the quantum soliton,
we need to know
$ -iN \logdetD $
explicitly for an arbitrary scalar field configuration.
One generally resorts to some local approximation,
such as the gradient expansion [\MWZ, \gradient],
accurate for slowly-varying configurations.
The gradient expansion, however,
breaks down for topological solitons
in the theories that we are considering.
Another approach to computing the fermion loop contribution
relies on the fact that for static solitons
$ (iN/T) \logdetD $
is just the energy of the ``Dirac sea,''
the sum of negative eigenvalues of the Dirac equation in
the soliton background~[\DHN].
Unfortunately,
the Dirac eigenvalues
must be numerically computed~[\KPR]
for each separate background considered,
rendering this approach inconvenient
for a variational problem.

In this paper, we propose a hybrid
of the gradient expansion and eigenvalue sum methods.
Following an idea of Wasson and Koonin [\WK],
we use the WKB approximation
to estimate the Dirac sea eigenvalues
for an arbitrary static scalar field background.
We then sum these to obtain
a local expression for the fermion vacuum energy.
Unlike the gradient expansion, this expression
is finite for topologically nontrivial configurations.
Using this WKB approximation,
we extremize the effective action
to find the form of the quantum soliton
in the large-$N$ theory.

We illustrate this method on a well-known example,
the kink of the \two--dimensional $\phi^4$ theory coupled to fermions.
The classical kink is reviewed in sect.~2.
In sect.~3,  we derive the WKB approximation for the
large-$N$ effective action in this theory.
This result is used in sect.~4 to find the form of the quantum kink,
which is contrasted to the classical kink.
The question of fermion stability is also discussed.
In sect.~5,
we present our conclusions
and discuss the features of the model
that we expect are generic to strongly-coupled solitons.

\chapter{Classical Kinks}

We begin by recalling the form and quantum numbers
of the classical kink [\solitons].
The \two--dimensional $\phi^4$ theory
coupled to $N$ flavors of fermion
has the Lagrangian
$$
\L
= \half \left(\del_\mu \phi \right)^2
- {\lam\over 4N} \left( \phi^2 - Nv^2 \right)^2
+ \sum_{i=1}^N \psiibar
               \left( i\delslash - {g \over \sqrt{N} } \phi \right)
               \psii.
\eqn\eelag
$$
The $N$-dependence of the parameters has been
chosen so that this theory has a sensible $N \to \infty$ limit.
If we rewrite $\phi$ as $\sqrt{N} \vphi$,
the parameter $N$ becomes an overall scale,
$$
\L
= N \left[ \tshalf \left(\del_\mu \vphi \right)^2
- \tsfourth  \lam \left( \vphi^2 - v^2 \right)^2 \right]
+ \sum_{i=1}^N \psiibar
               \left( i\delslash - g \vphi \right)
               \psii.
\eqn\eeNlag
$$
In the vacuum state  $|\vphi| = v$,
the scalar field has mass $\sqrt{2\lam} v$
and the fermion field mass $gv$.
In two dimensions, $v$ is dimensionless,
the scalar self-coupling $\lam$ has dimension 2,
and the Yukawa coupling $g$ dimension 1.
It is convenient to substitute
for $\lam$ and $g$ the parameters
$$
\xcl  = \sqrt{2 \over \lam v^2},
\qquad
y = g \sqrt{2 \over \lam}.
\eqn\eeparam
$$
The parameter $\xcl$ is
proportional to the scalar field Compton wavelength
(and, as we will see, the size of the classical kink),
and will serve as the overall scale of length and energy
in the theory.
There are two dimensionless parameters,
$v$ and $y$,
the latter being
proportional to the ratio of fermion and scalar masses.

The Lagrangian \eeNlag~gives rise to the field equations
$$
\eqalignno{
\del^2 \vphi + \lam \vphi^3 - \lam v^2 \vphi
& = - g {1\over N} \sum_{i=1}^N \psiibar \psii,
& \eqname\eescalareq
\cr
\left( i\delslash - g\vphi \right) \psii
& = 0.
& \eqname\eediraceq
\cr}
$$
The topologically nontrivial solutions of these equations
give a ``classical'' description
of the soliton states in the Hilbert space,
which is accurate when the quantum corrections
are small.
If we neglect the fermion source term,
the scalar field equation \eescalareq~has the
well-known static kink solution
$$
\vphicl(x) = v ~ \tanh \left( x\over \xcl \right),
\eqn\eeclasskink
$$
which is the lowest energy state with topological charge
$ [\vphi(\infty)-\vphi(-\infty)]/ 2v = 1$.
There is also an anti-kink solution
which interpolates from $v$ to $-v$,
with topological charge $-1$.
The Dirac equation \eediraceq~
in the kink background \eeclasskink~
has a self-conjugate zero mode solution
$$
\psi_0(x)  = \left( [ \sech(x/\xcl) ]^y \atop 0  \right),
\qquad {  \gamma^0 = \sigma_1, \atop \gamma^1  = i \sigma_3.}
\eqn\eezeromode
$$
The state with the zero mode occupied
has the same energy as that with the zero mode unoccupied.
Since there is a zero mode for each flavor $i$,
the kink is $2^N$-fold degenerate.
If $n$ of the zero modes are occupied,
the kink has fermion number $n-\half N$,
which ranges from $-\half N$ to $\half N$ [\JR].
The anti-kink too has degeneracy $2^N$,
and fermion number ranging from $-\half N$ to $\half N$.
Although the fermion zero modes
increase the degeneracy of the kink,
their contribution to the source term in
the scalar field equation \eescalareq~vanishes,
so the kink \eeclasskink~ remains a solution
even in the presence of fermions.
The energy of the classical kink
$$
E \bkphicl  = {2 \sqrt{2} \over 3} N \sqrt{\lam} v^3
  = {4\over 3} v^2 \scale
\eqn\eeclasskinkenergy
$$
has no dependence on the Yukawa coupling $g$,
even though the kink carries fermion number
due to the zero modes.

This classical picture leads to the fascinating possibility that,
even if fermion number is conserved,
``ordinary'' fermions may be unstable to the formation of
solitons carrying  fermion number.
A configuration
consisting of a widely-separated kink and anti-kink,
each carrying fermion number $\half N$,
has zero topological charge, fermion number $N$,
and energy
$  {8\over 3} v^2 \tscale $.
On the other hand,
a set of $N$ widely-separated fermions
in the vacuum background $\vphi=v$,
a state which has the same quantum numbers,
has  energy $Ngv= y \tscale $.
Thus, when $y > {8\over 3}v^2$,
it is energetically favorable for a state of $N$ fermions
to coalesce onto a spontaneously created kink/anti-kink pair.
Each kink acts as a kind of bound state of $\half  N$ fermions.
Even more surprising,
in a theory with one flavor of fermion ($N=1$),
a single fermion could split into a kink/anti-kink pair,
each with fermion number $\half$.

This putative instability occurs only when $y$ is large,
however,
where the quantum corrections from fermion loops are important
and the semi-classical approximation breaks down.
To determine whether fermions are truly unstable,
one must compare their energy not with that of a classical kink,
but of a ``quantum kink,''
which includes the effects of quantum corrections.
The quantum kink extremizes
not the action but rather the effective action.
In the next section we will derive a local
expression for the effective action
suitable for finding the quantum kink.

\chapter{Effective Action for Kinks}

Quantum solitons are field configurations
that extremize the effective action,
which includes quantum corrections.
To find the form of quantum solitons,
one needs an explicit local expression
for the effective action.
The familiar gradient expansion,
however, diverges for topologically nontrivial
configurations in \two--dimensional
$\phi^4$ theory.
In this section,
we derive an alternative local approximation
for the effective action
that is finite for kinks.

Since we are interested in the properties of solitons
for large Yukawa coupling $g$,
the effective action must be
calculated nonperturbatively in $g$.
This can be done by taking
the number $N$ of fermion flavors to be large,
holding $\lam$, $v$, and $g$ fixed,
and calculating to leading order in $1/N$.
Scalar field fluctuations are subleading in $1/N$,
so only fermion loops contribute to the large-$N$ effective action
$$
\eqalign
{
\Seff \bkphi
& = \int \d^2 x~\Leff (\vphi)
\cr
& = N \int \d^2 x \left[ \tshalf \left(\del_\mu \vphi \right)^2
- \tsfourth \lam \left( \vphi^2 - v^2 \right)^2 \right]
 ~~-~~ iN \logdet
\cr
&~~~~~~ +   \int \d^2 x~\dL (\vphi) ~~+ ~~iN \logdetv.
\cr
}
\eqn\eeeffact
$$
We have added the counterterm
$$
\dL (\vphi) = A N  (\vphi^2 - v^2)
\eqn\eeAcounterterm
$$
to tame the divergent contributions
of the fermion determinant
to the one- and two-point functions,
and the overall constant
$ iN \logdetv $
to ensure that $ \Seff [\vphi = v] = 0$.
The coefficient $A$ is fixed by
requiring the one-point function to vanish at $\vphi=v$,
$$
0 = 2Av + ig \int {\d^2 p \over (2\pi)^2 }
     ~\tr \left( i \over \pslash - gv \right),
\eqn\eeonepoint
$$
so that $v$ remains the minimum of the effective potential.
With a cutoff $\Lambda$ on the spatial momentum $\pone$,
eq.~\eeonepoint~ gives
$$
\dL (\vphi) = -{Ng^2 \over 2\pi} (\vphi^2-v^2)
\int_0^\Lambda {\d \pone \over \sqrt{\pone^2 + g^2 v^2} }.
\eqn\eecounterterm
$$
This counterterm also renders finite the two-point function
$$
    \Gamtwo (p)\Big|_{p=0} = - 2\lam v^2 - {g^2\over\pi},
\eqn\eetwopoint
$$
where $\sigma=\vphi - v$.
Fermion loop contributions to all other Green functions are finite.

We must write the effective action \eeeffact~in
a more tractable form
if we are to find the quantum kink explicitly.
The gradient expansion [\MWZ, \gradient]
$$
   \Leff (\vphi)
  = - V_\eff (\vphi) + \Ltwo (\vphi) + \cdots
\eqn\eederivexp
$$
is a useful approximation for slowly-varying fields.
The first term in this expansion is minus the effective potential
$$
          { V_{\eff} (\vphi) \over N}
=  {\lam \over 4} \left( \vphi^2 -  v^2 \right)^2
  + { g^2 \over 4\pi } \vphi^2 \ln \left( \vphi^2 \over v^2 \right)
  - { g^2 \over 4\pi } \left( \vphi^2 - v^2 \right).
\eqn\eeeffpot
$$
The term with two derivatives is
$$
   {\Ltwo (\vphi)   \over N}
=  \half \left[ 1 + {1  \over 12\pi\vphi^2} \right]
   \left( \del_\mu \vphi \right)^2.
\eqn\eetwoderiv
$$
At this point,
we discover that the gradient expansion fails
for topological solitons in this theory;
any configuration $\vphi(x)$
with unit topological charge
must pass through $\vphi=0$ somewhere,
at which point $\Ltwo (\vphi)$,
as well as higher order terms,
diverges.
This failure is quite general.
For the gradient expansion to converge,
field gradients must be small relative to $ g\vphi $,
the ``local fermion mass.''
Since the latter vanishes
at the core of solitons
with fermion zero modes,
the gradient expansion necessarily breaks down there,
no matter how slowly varying the field.

An alternative approach
for a static scalar field background such as the kink
is to express the effective action
in terms of Dirac equation eigenvalues [\DHN,\CL].
For time-independent $\vphi (x) $,
the effective action equals
$ - \Eeff \bkphi T$,
where $ T = \int \d t$ and $ \Eeff \bkphi $
is the energy of the configuration,
$$
\Eeff \bkphi = \Ecl \bkphi + \Q \bkphi,
\eqn\eeeffenergy
$$
a sum of the classical energy
$$
\Ecl \bkphi
= {N \over \xcl}
 \int_{-\infty}^{\infty}
 \d z \left[  \half \left(\d \vphi \over \d z \right)^2
+  {1 \over 2v^2} \left( \vphi^2 - v^2 \right)^2 \right],
\qquad z= {x\over\xcl},
\eqn\eeclassenergy
$$
and the quantum correction, the fermion vacuum energy,
$$
\Q \bkphi
= {iN\over T} \logdet - {iN\over T} \logdetv + \dE \bkphi.
\eqn\eequantumcorr
$$
The first term in eq.~\eequantumcorr~can
be interpreted as the energy of the Dirac sea
in the background $\vphi(x)$.

To  write eq.~\eequantumcorr~more explicitly,
we observe that
the Dirac equation \eediraceq~ implies
that the spinor components
$ \psii = \left(\psii_+ \atop \psii_- \right) $
obey the Schr\"odinger-type equations
$$
\left[  {\d^2 \over \d z^2} - y^2 V_\pm (z) - y^2
+ \xcl^2 \eps^2_\pm \right] \psi_\pm= 0
\eqn\eeschrod
$$
in a static background $\vphi(z)$,
where
$$
V_\pm (z) = \left( {\vphi^2 \over v^2} - 1 \right)
\mp {1\over yv} {\d \vphi \over \d z}.
\eqn\eeschrodpot
$$
We restrict $\vphi(z)$ to
configurations of unit topological charge that obey
$\vphi(-z) = -\vphi(z)$;
the Schr\"odinger potentials $\Vsig (z)$ are then even,
and the solutions $\psisig (z)$ can be taken to be parity eigenstates.
(Here $\sig = \pm$ labels the upper and lower spinor components.)
Since $\bigl| \phi(\pm\infty) \bigr| = v$,
the potentials $\Vsig (z)$ vanish at $\pm \infty$,
so eq.~\eeschrod~has a continuous spectrum of states
classified by their asymptotic momentum,
$ k = \sqrt{ \xcl^2 \eps^2 - y^2} $,
and their parity.
The asymptotic forms of the continuum wavefunctions
$$
\eqalign
{
\psievensig (k,z)
&  \mathrel{\mathop{\longrightarrow}\limits_{z \to \pm \infty}}
  \cos( kz \pm \tshalf \devensig (k) ),
\crr
\psioddsig (k,z)
&   \mathrel{\mathop{\longrightarrow}\limits_{z \to \pm \infty}}
   \sin( kz \pm \tshalf \doddsig (k) ),
\cr}
\eqn\eephaseshift
$$
serve to define the phase shifts
$\devensig (k)$  $\left[\doddsig (k)\right]$
for the
even (odd) parity states.
If we put the system into a box,
$|z| \le \half L$,
with periodic boundary conditions,
eq.~\eephaseshift~implies that the allowed momenta satisfy
$ k_{\sig n} L + \deltasig (k) = 2\pi n$.
Eq.~\eeschrod~may also have a series of discrete bound states with eigenvalues
$ \eps_{\sig i}^2 <  y^2/\xcl^2$.
Because the configuration has topological charge 1,
the upper spinor component is guaranteed [\JR] to have a zero mode,
$\eps_{+} = 0$.
Thus,
any configuration
with unit topological charge
can carry fermion quantum numbers.

The difference of fermion loop contributions can be written
as the shift of the Dirac sea energy [\DHN]
$$
{i N\over T}  \logdet - {i N\over T} \logdetv
= - ~ \tshalf  N \sum_{\sig} \sum_{\lam}
   \left( \epssiglam - \epssiglam^\zo \right) ,
\eqn\eevacenergy
$$
where $\epssiglam$ denotes the positive root of $\epssiglam^2$,
and $\epssiglam^\zo$ are the Dirac eigenvalues in
the constant configuration $\vphi (x) = v$.
Eq.~\eevacenergy~may be separated into the
sum over discrete eigenvalues
$$
\Edis \bkphi = ~-~ \tshalf N \sum_{\sig} \sum_i
  \left( \eps_{\sig i} - {y \over \xcl}  \right)
\eqn\eediscenergy
$$
and the sum over continuum eigenvalues
$$
\Econ \bkphi
=~-~\tshalf N \sum_{\sig} \sum_{n>0} \sum_{\rm parity}
            \left[  \eps (k_{\sig n}) - \eps (k_{\sig n}^\zo) \right],
\qquad \eps (k) = { \sqrt{k^2 + y^2}  \over \xcl}.
\eqn\eecontenergy
$$
Using
$ k_{\sig n} L + \deltasig (k) = k_{\sig n}^\zo  L = 2\pi n$,
and letting $L \to \infty$,
we can write eq.~\eecontenergy~as [\DHN]
$$
\Econ \bkphi
= N \sum_{\sig } \int_{0}^{\Lambda}
           {\d k \over 4\pi} {\d \eps \over \d k}
    \left[ \devensig (k)  + \doddsig (k) \right]
= N \sum_{\sig } \int_{0}^{\Lambda}
           {\d k \over 2\pi} {\d \eps \over \d k} \delta_{\sig} (k) ,
\eqn\eephaseintegral
$$
where
$ \delta = \half \left( \deven + \dodd \right)$.
The integral over $k$ diverges as the momentum cutoff $\Lambda$ is removed,
but this divergence is cancelled by the counterterm energy
$$
\dE \bkphi = - \int_{-\infty}^{\infty} \d z ~\dL (\vphi)
= {y^2 \over 2\pi} \scale \int_0^\Lambda {\d k \over \sqrt{k^2 + y^2} }
\int_{-\infty}^\infty \d z \left( {\vphi^2\over v^2} - 1 \right).
\eqn\eeenergycounterterm
$$
The sum of eqs.~\eediscenergy, \eephaseintegral, and \eeenergycounterterm,
$$
\Q \bkphi = \Edis \bkphi + \Econ \bkphi + \dE \bkphi
\eqn\eequantumsum
$$
is precisely the fermion vacuum energy \eequantumcorr.

The expression \eequantumsum~for the fermion vacuum energy is
much more explicit than eq.~\eeeffact,
and can even be computed analytically for certain scalar field
configurations~[\CY].
For an arbitrary background, however,
$\eps_{\sig i}$ and $\delta_{\sig} (k)$  must be
computed numerically~[\KPR].
Wasson and Koonin [\WK]
showed how to speed up the convergence
of these ``brute force'' numerical calculations
by employing the WKB approximation
for the high momentum phase shifts,
but the discrete eigenvalues and low momentum phase shifts
must still be computed numerically for each separate field
configuration.
Thus, eq.~\eequantumsum~is still not very convenient
for extremizing the effective action.\foot{
Campbell and Liao [\CL]
were able to extremize
{}~\eeeffenergy~using powerful inverse scattering methods,
but only for the special case $y=1$.
}

Taking our cue from ref.~[\WK],
we adopt the WKB approximation
for all the Dirac eigenvalues,
both continuous and discrete,
and use them in eq.~\eequantumsum~to
obtain a local expression for the energy of
an arbitrary scalar field configuration.
The resulting expression will be accurate
for field configurations slowly varying on
the scale of the fermion Compton wavelength,
but unlike the gradient expansion,
does not diverge for solitons.
We will then use this approximate expression
to find the form of the quantum kink
in sect.~4.

In the WKB approximation,
the continuum eigenfunctions of eq.~\eeschrod~are
$$
\eqalign
{
\psievensig^\wkb  (k,z)
& = {1 \over \sqrt{\ksig (z)}} \cos
\left( \textstyle \int_0^z \ksig (z^\prime) \d z^\prime \right),
\cr
\psioddsig^\wkb (k,z)
& = {1 \over \sqrt{\ksig (z)}} \sin
\left( \textstyle \int_0^z \ksig (z^\prime) \d z^\prime \right),
\qquad \ksig (z) = \sqrt{ k^2 - y^2 \Vsig (z)},
\cr
}
\eqn\eewkbwavefunction
$$
whence the phase shift defined through eq.~\eephaseshift~is given by
$$
\deltasig^\wkb (k) = \int_{-\infty}^{\infty} \d z \bigl[ \ksig (z) - k \bigr],
\eqn\eewkbphaseshift
$$
independent of parity.
(We assume $ \Vsig (z) \le 0 $ everywhere;
this will be true if $\vphi (z) $ does not vary too rapidly.)
Using the WKB phase shifts
\eewkbphaseshift~in the integral \eephaseintegral~and
adding the counterterm energy \eeenergycounterterm,
we find
$$
\eqalign{
\Econ^\wkb \bkphi + \dE \bkphi
&
= {y^2\over 4\pi} \scale \int_{-\infty}^{\infty}
\d z \bigg[  \left(1 -  {\vphi^2 \over v^2} \right)
-  \left( \sqrt{-V_+} + \sqrt{-V_-} \right)
\cr
&
\quad + (1 + V_+) \log \left( 1 + \sqrt{-V_+} \right)
+ (1 + V_-) \log \left( 1 + \sqrt{-V_-} \right) \bigg].
\cr
}
\eqn\eewkbcontenergy
$$
We also need to approximate the sum over
discrete eigenvalues \eediscenergy.
In the WKB approximation,
the Schr\"odinger equation
\eeschrod~has discrete eigenvalues $\eps$
whenever $\wsig (\eps)$,
defined by
$$
\wsig (\eps)
= {1\over \pi} \int_{-\infty}^\infty \d z~\ksig(z) \Theta (\ksig^2(z)),
\qquad \ksig (z) = \sqrt{ \xcl^2 \eps^2 - y^2 - y^2 \Vsig (z)},
\eqn\eebohrsommerfeld
$$
equals half an odd integer, $w \in \ZZ + \half$.
The number of discrete eigenstates is
given by the integer closest to $\wsig (y/\xcl)$.
We define $ \epssig (w) $
by inverting eq.~\eebohrsommerfeld~and
setting $\epssig(w) = 0$ for $0 \le w \le \wsig (0)$.
The sum over discrete eigenvalues \eediscenergy~ in the WKB approximation
is then written
$$
\Edis^\wkb \bkphi = ~-~ \tshalf \sum_{\sig}
\sum_{ {w\in\ZZ + \half} \atop {0 < w\le \wsig(y/\xcl) } }
  \left( \eps_{\sig} (w)  - {y \over \xcl}  \right).
\eqn\eewkbdiscenergy
$$
We separate this into two terms
$$
\Edis^\wkb \bkphi
= \Edisone \bkphi + \Edistwo \bkphi,
\eqn\eewkbsplit
$$
where $\Edisone \bkphi$ is the integral approximation of the sum
\eewkbdiscenergy
$$
\Edisone \bkphi
 = ~-~ \tshalf \sum_{\sig} \int_0^{\wsig (y/\xcl)}
   \d w \left( \epssig (w)  - {y \over \xcl}  \right),
\eqn\eeintegralapprox
$$
and $\Edistwo \bkphi$ is the remainder.
The integral \eeintegralapprox~may be rewritten
$$
\eqalign{
\Edisone \bkphi
& = ~-~ \tshalf \sum_{\sig} \int_0^{y/\xcl}
   \d \eps ~ \wsig (\eps)
\cr
& = ~-~ {1 \over 2 \pi} \int_{-\infty}^\infty \d z ~\sum_{\sig}
    \int_0^{y/\xcl}  \d \eps ~\ksig(z) \Theta (\ksig^2(z))
\cr
& = {y^2\over 4\pi } \scale  \int_{-\infty}^{\infty}
\d z \bigg[
  \left( \sqrt{-V_+} + \sqrt{-V_-} \right)
+ (1 + V_+) \log
\left( \sqrt{ \left| 1 + V_+ \right|} \over 1 + \sqrt{-V_+} \right)
\cr
&\qquad\qquad\qquad\qquad\qquad\qquad\qquad\qquad\qquad
+ (1 + V_-) \log
\left( \sqrt{ \left| 1 + V_- \right|} \over  1 + \sqrt{-V_-} \right)\bigg].
\cr
}
\eqn\eewkbintegral
$$
Adding the contributions from the continuum \eewkbcontenergy~and
discrete \eewkbsplit~states,
we obtain
$$
\eqalign{
\QWKB \bkphi
= {y^2\over 4\pi} \scale \int_{-\infty}^{\infty}
\d z
&
\bigg[
\left( 1 - {\vphi^2\over v^2}\right)
+ \half \left(  {\vphi^2 \over v^2} + {1\over yv} {\d \vphi \over \d z} \right)
  \log \left|   {\vphi^2 \over v^2} + {1\over yv} {\d \vphi \over \d z} \right|
\cr
&
+ \half \left(  {\vphi^2 \over v^2} - {1\over yv} {\d \vphi  \over \d z}
\right)
  \log \left|   {\vphi^2 \over v^2} - {1\over yv} {\d \vphi  \over \d z}
\right|
\bigg] + \Edistwo \bkphi
\cr
}
\eqn\eewkbquantumcorr
$$
for the fermion vacuum energy in the WKB approximation.

When $\vphi (z)$ is slowly varying
on the scale of the fermion Compton wavelength,
the number of discrete states $\wsig (y/\xcl)$ is large,
the sum \eewkbdiscenergy~is well approximated by the
integral \eeintegralapprox,
and $\Edistwo \bkphi$
is much smaller than $\Edisone \bkphi$.
If we therefore neglect $\Edistwo \bkphi$,
eq.~\eewkbquantumcorr~
provides a completely explicit local expression for
the energy of a static configuration
$$
\eqalign{
\Eeff^\wkb  \bkphi
&  = {N \over \xcl}
\int_{-\infty}^{\infty}
\d z \left[  \half \left(\d \vphi \over \d z \right)^2
+  {1 \over 2v^2} \left( \vphi^2 - v^2 \right)^2 \right]
\cr
& + {y^2\over 4\pi} \scale \int_{-\infty}^{\infty}
\d z \bigg[
\left( 1 - {\vphi^2\over v^2}\right)
{}~+ \half \left(  {\vphi^2 \over v^2} + {1\over yv} {\d \vphi \over \d z}
\right)
  \log \left|   {\vphi^2 \over v^2} + {1\over yv} {\d \vphi \over \d z} \right|
\cr
&\qquad\qquad\qquad\qquad\qquad\qquad\qquad
+ \half \left( {\vphi^2 \over v^2} - {1\over yv} {\d \vphi  \over \d z} \right)
  \log \left|  {\vphi^2 \over v^2} - {1\over yv} {\d \vphi  \over \d z} \right|
\bigg].
\cr
}
\eqn\eelocaleffenergy
$$
For $\vphi (z)$ constant,
eq.~\eelocaleffenergy~reduces
to the effective potential \eeeffpot.
When $\vphi (z)$ is not constant,
eq.~\eelocaleffenergy~
yields a correction to the effective potential
which,
unlike the gradient expansion,
does not diverge for configurations
going through $\vphi=0$.

We conclude this section by comparing
the WKB approximation of the fermion vacuum energy
of the classical kink,
$\vphicl(z) = v~\tanh(z)$,
with the known exact result.
The WKB approximation should be accurate for $y \gg 1$,
when $\vphicl(z) $ is slowly-varying relative to the
fermion Compton wavelength.
The Dirac equation can be solved analytically
in the classical kink background.
Using the resulting eigenvalues,
Chang and Yan [\CY]
computed the exact fermion loop correction
\eequantumsum~to the energy of the classical kink
$$
\Q \bkphicl = \scale \Delt (y) .
\eqn\eechangyanQ
$$
The function $\Delt (y)$ is given by a complicated integral,
but for integer $y$ it simplifies to [\CY]
$$
\Delt (y)
 ~=~
{y^2\over \pi} ~+~ \sum_{n=1}^{y-1}
\left( - \sqrt{2yn-n^2}
      ~+ ~{2\over \pi} \sqrt{y^2-n^2} \arctan \sqrt{ {y^2\over n^2} - 1 }
        ~\right),
\qquad y \in \ZZ.
\eqn\eechangyan
$$
Using the Euler-Maclaurin formula,
we obtain the large $y$ behavior of eq.~\eechangyan
$$
\Delt (y)
=  \left(  {3 \over 2\pi} - {\pi \over 8}  \right) y^2
 + \left(  2 \sqrt{2} \beta + {1 \over 3\sqrt{2} } \right) \sqrt{y}
 + O(1),
\qquad
\beta
= \sum_{k=1}^\infty  { (4k-5)!! \over 2^{2k}  (2k)! } B_{2k} \approx
.0206\ldots
\eqn\eechangyanapprox
$$
where $B_{2k}$ are the Bernoulli numbers.
The series defining $\beta$ is asymptotic,
so we only keep 4 or 5 terms in the sum.

The WKB approximation is obtained by substituting
$\vphicl(z)$
into eq.~\eewkbquantumcorr~and expanding for large $y$
$$
\QWKB \bkphicl = {N\over \xcl}
\left[ \left(  {3 \over 2\pi} - {\pi \over 8} \right) y^2
      + {1 \over 6} \sqrt{y} + O(1) \right]
+ \Edistwo \bkphicl.
\eqn\eewkbkinkcorr
$$
Using the WKB approximation for the discrete eigenvalues
$ \eps_{\sig i} $ together with the Euler-Maclaurin formula,
we find the leading behavior of the remainder term
$$
\eqalign{
\Edistwo \bkphicl
&
= {N\over\xcl}
\left[  \left( 2 \sqrt{2} \beta + {1\over 3\sqrt{2}} - {1\over 6} \right)
        \sqrt{y} + O(1) \right].
\cr
}
\eqn\eeeulermaclaurin
$$
Thus the WKB approximation of the fermion vacuum energy is
$$
\QWKB \bkphicl
= {N\over \xcl}
\left[  \left(  {3 \over 2\pi} - {\pi \over 8}  \right) y^2
      + \left(  2 \sqrt{2} \beta + {1 \over 3\sqrt{2} } \right) \sqrt{y}
      + O(1) \right],
\eqn\eewkbeulerkink
$$
in agreement with eq.~\eechangyanapprox~to this accuracy.
The $y^2$ term, of course,
is just the contribution from the effective potential \eeeffpot.
The non-analytic subleading $\sqrt{y}$ dependence
cannot be seen in the gradient expansion,
but is correctly given by the
WKB approximation \eewkbquantumcorr.

Obviously,
the coefficient obtained for the subleading $\sqrt{y}$ dependence
would be incorrect if we made the further approximation
of dropping $\Edistwo \bkphi$,
as was done in obtaining the local expression \eelocaleffenergy.
Nonetheless,
eq.~\eelocaleffenergy~correctly gives
the order of the subleading dependence.
In general, it provides a useful estimate
of the correction to the effective potential
for a spatially-varying field.

\chapter{Quantum Kinks}

The quantum kink extremizes the effective action
of the \two--dimensional $\phi^4$ theory.
For small Yukawa coupling $y$,
the effective action \eeeffact~differs
only slightly from the classical action,
so the quantum kink nearly
coincides with the classical kink.
When $y$ is large, however,
fermion loop corrections are important,
and the quantum kink differs significantly
from the classical kink.

To find the explicit form of the quantum kink,
we use the local approximation \eelocaleffenergy~
for the energy  $\Eeff \bkphi$
of a static scalar field configuration
derived in sect.~3.
The equation of motion for the quantum kink
follows from extremizing eq.~\eelocaleffenergy,
$$
\eqalign{
\left( 1 + {1\over 4\pi v^4} {\vphi^2 \over f(\vphi) } \right)
{\d^2 \vphi \over \d z^2}
&= 2 \vphi \left( {\vphi^2 \over v^2}  - 1 \right)
+ {y^2 \over 4\pi v^2} \vphi \log \bigl| f(\vphi) \bigr|
+ { 1\over 2 \pi v^4} {\vphi \over f(\vphi)}
  \left( \d\vphi \over \d z \right)^2 ,
\crr
f(\vphi)
& = {\vphi^4 \over v^4} - {1\over y^2 v^2}
  \left( \d\vphi \over \d z \right)^2 .
\cr}
\eqn\eequantumeom
$$
Using the program COLSYS [\COLSYS],
we have solved this equation numerically
for various values of the parameters
subject to the boundary condition
$\vphi(\pm \infty) = \pm v$.
The solutions obtained
interpolate smoothly between $-v$ and $v$.
Indeed,
their profiles are almost indistinguishable from
the hyperbolic tangent shape of the classical kink (see fig.~1).
The slope of the quantum kink differs
from that of the classical kink, however,
being much steeper for certain values of the parameters.

We can more easily see how the slope of the quantum kink
depends on the parameters of the theory
by restricting $\vphi (x)$ to the one-parameter family of functions
$$
\vphi_{\xqu}  (x) = v ~\tanh \left(x\over \xqu \right),
\eqn\eeansatz
$$
where
$ \xqu $
is the ``size'' of the ansatz.
We write the energy of the ansatz
$$
\Eeff (\zqu) = \Ecl (\zqu) + Q (\zqu),
\qquad \zqu = {\xqu \over  \xcl},
\eqn\eeansatzenergy
$$
where
$ \zqu $
is the ratio of the size of the ansatz
to that of the classical kink.
The classical contribution
$$
\Ecl (\zqu)  = {N\over \xcl}
\left[ {2\over 3} v^2 \left( \zqu + {1\over\zqu} \right) \right]
\eqn\eeansatzclassical
$$
has a minimum at $\zqu = 1$, of course.
The quantum contribution
is obtained by substituting the ansatz \eeansatz~
into the WKB approximation \eewkbquantumcorr~
and retaining the leading power of $y$
$$
Q^\wkb (\zqu) =
{N\over \xcl}
\left[
     \left( {3\over 2\pi}-{\pi\over 8} \right) y^2 \zqu
 + O \left( \sqrt{ y\over \zqu}        \right)
\right].
\eqn\eeansatzwkb
$$
The WKB approximation is accurate when the
neglected terms are small,
which requires $ \zqu \gg 1/y $.
That is,
the size of the ansatz must be much larger
than the  fermion Compton wavelength ($ \xqu \gg 1/gv$).

In the following discussion,
we assume large Yukawa coupling $y \gg 1$.
The size of the quantum kink is found by
minimizing \eeansatzenergy,
$$
\zqu =
  \left[
 1    + \left( {9\over 4\pi}-{3 \pi\over 16} \right) {y^2\over v^2}
  \right]^{-1 / 2} ,
\eqn\eekinksize
$$
and depends on the values of
both dimensionless parameters $y$ and $v$.
When $ v \gg y $, the kink size $\zqu \approx 1$
and the quantum kink reduces to the classical kink,
because the classical contribution to the energy
is dominant in this regime.
On the other hand,
when $v \ll y$ (but $v \gg 1$),
the kink size
$\zqu  \approx
\left( {9\over 4\pi} - {3\pi\over 16}\right)^{-{1\over 2}} (v/y)
\approx 2.8(v/y)$;
the quantum kink is much smaller than the classical kink.
The energy of the quantum kink in this regime,
$ \Eeff
\approx \left( {4\over \pi} - {\pi\over 3}\right)^{1 \over 2} vy \tscale
\approx .48 \, vy \tscale $,
is larger than the classical kink energy
\eeclasskinkenergy~
due to the positive fermion vacuum energy \eeansatzwkb.

When $v \lsim 1$,
the WKB approximation \eeansatzwkb~breaks down
because the kink size is no longer much larger than the fermion
Compton wavelength.
By using the exact Dirac eigenvalues
for the background \eeansatz~rather than
the WKB eigenvalues,
however,
we can calculate the fermion vacuum energy
$Q(\zqu)$ without approximation,
just as for the classical kink.
We find
$$
Q(\zqu) = \scale y \, \Deltt (y\zqu),
\qquad
\Deltt (t) = {\Delt(t) \over t },
\eqn\eeansatzquantumcorr
$$
where $\Delt(y)$ is the fermion vacuum energy of the classical kink
defined in sect.~3.
The function $\Deltt(t)$ is shown in fig.~2,
and equals $1/\pi$ at its minimum $t = 1$.
(That its minimum is at $t=1$ can be seen from
eq.~\eechangyan~and from
$$
{\d \Delt \over \d t}
 ~=~ {1\over 2} \delta_{t 0}
 ~+~ {t \over \pi}
 ~+~ \sum_{n=1}^{t-1}
\left(
      - \sqrt{ n \over 2t - n}
      ~+~  {2\over \pi} \sqrt{t - n \over t + n }
          \arctan \sqrt{ {t^2\over n^2} - 1 } ~
\right),
\quad t  \in \ZZ,
\eqn\eedeltaprime
$$
obtained by a calculation similar to that in ref.~[\CY].)
When $v \ll 1$ (and $y\gg 1$),
the fermion vacuum contribution \eeansatzquantumcorr~dominates the energy,
so the size of the quantum kink
is determined by the minimum of $\Q (\zqu)$,
that is, $\zqu \approx 1/y$.
The quantum kink energy,
$ \Eeff \approx (y/\pi) (N/\xcl)$,
is much larger than that of the classical kink \eeclasskinkenergy.

The classical contribution to the energy $\Ecl(\zqu)$
is minimized for $\zqu=1$,
when the ansatz size equals the scalar field Compton wavelength,
$\xqu = \xcl$.
The quantum contribution $Q(\zqu)$
is minimized for $\zqu = 1/y$,
when the ansatz size equals the fermion Compton wavelength,
$\xqu = \xcl/y = 1/gv$.
The size of the quantum kink always lies somewhere
between these two values.
The three limits we considered above
$$
\eqalign{
&
1 \ll y \ll v
{}~~\Rightarrow~~
\zqu \approx 1,
\qquad \qquad \qquad \qquad \qquad
\Eeff \approx {4v^2\over 3} \scale,
\cr
&
1 \ll v \ll y
{}~~\Rightarrow~~
\zqu \approx
\left( {9\over 4\pi} - {3\pi\over 16} \right)^{-1/2} {v\over y},
\qquad ~\,
\Eeff  \approx \left( {4\over \pi} - {\pi\over 3} \right)^{1/2} vy \scale,
\cr
&
v \ll 1 \ll y
{}~~\Rightarrow~~
\zqu \approx {1\over y},
\qquad \qquad \qquad \qquad \qquad
\Eeff  \approx {y\over \pi} \scale.
\cr
}
\eqn\eesummary
$$
correspond to the regime in which the classical energy is dominant
($v\gg y$),
the regime in which the fermion vacuum energy is dominant
($v\ll 1$),
and the regime in which both contributions are important
($1\ll v\ll y$).
For $ v \gsim y$,
the classical and quantum kink nearly coincide,
while for $ v \lsim y$,
the quantum kink is smaller and has greater energy than
the classical kink.
Note that due to the fermion vacuum contribution \eeansatzquantumcorr,
the energy of the kink is bounded below by
$ (y/\pi) (N/\xcl) = Ngv /\pi $,
that is, $1/\pi$ times the mass of $N$ fermions.

We now turn to the question of fermion stability.
In sect.~2, we saw that for sufficiently strong Yukawa coupling,
$ y > {8\over 3} v^2 $,
a state of $N$ widely-separated fermions
has greater energy than a kink/anti-kink pair,
computed in the classical approximation,
so one might expect
a kink/anti-kink pair to appear spontaneously,
with the fermions coalescing to occupy the zero modes.
Since the zero modes do not increase the kink energy,
the energy of the fermions on the kinks is independent
of $y$ in the classical approximation,
and would be much less than the energy of
the fermions in a constant scalar field background
for $ y \gg v^2$.
The kink binding energy could approach 100\%
for very large Yukawa coupling.

Instead we have found that,
for large $y$,
quantum corrections significantly increase
the energy of the kink.
For $ v \gsim 1 $ (but $v \ll y$),
a kink/anti-kink pair has energy
$ \sim yv (N/\xcl)$,
greater than the energy of $N$ fermions,
so the fermions are stable.
For $v \lsim 1$,
the energy of a kink/anti-kink pair may be less than
$ y (N/\xcl)= Ngv$,
in which case a state of $N$ fermions
may be unstable to the formation of a kink
and anti-kink, each carrying fermion number $\half N$.
Since the kink energy is never less than
$ (y/\pi) (N/\xcl) $,
however,
the energy of a widely-separated kink/anti-kink pair
is not significantly less than that of the original fermions;
the binding energy per fermion cannot exceed
$ 1 - {2\over\pi} \sim  36\%$.

Up to this point, we have been chiefly concerned
with large Yukawa coupling, $y \gg 1$;
we conclude this section by briefly considering $ y \lsim 1$.
When $y$ is not large,
the WKB approximation is no longer useful,
but we can use the exact solution \eeansatzquantumcorr~
for the ansatz~\eeansatz.
The case $y=1$ is interesting,
because then $\zqu = 1$
minimizes both the classical and quantum contributions
to the energy;
the classical kink is an extremum of the effective
action restricted to the subspace of functions \eeansatz.
One might suspect from this that
the classical kink extremizes the effective action
over the space of all functions.
Campbell and Liao [\CL]
proved this to be the case
by using inverse scattering methods
(which were tractable only when $y=1$).\foot{
The stationary phase approximation of ref.~[\CL]
is equivalent to our large-$N$ approximation.}
Thus, the quantum kink exactly
coincides with the classical kink (for all values of $v$)
when $y=1$.\foot{
Interestingly, the theory is supersymmetric precisely
when $y=1$ [\CL, \super].}
As we have seen, they differ when $y \neq 1$.

The energy of the kink when $y=1$ is
$$
\Eeff  = \left( {4\over 3}v^2 + {1 \over \pi} \right) {N\over\xcl} ,
\qquad \qquad y=1,
\eqn\eecampbellliao
$$
so a kink/anti-kink pair will have less energy than $N$
widely-separated fermions when
$ v <  \sqrt{ {3\over 8} \left( 1 - {2\over\pi} \right) } \approx .37$.
It turns out that $N$ fermions are unstable to kink formation
only if $v < \sqrt{ 1\over {4\pi} } \approx .28$;
the most energetically favorable configuration of $N$ fermions
for $v > \sqrt{ 1\over {4\pi} }$
is a bag [\CL]. (See also ref.~[\MP].)

Finally, for small Yukawa coupling, $y < 1$,
the fermion Compton wavelength is larger than
the scalar field Compton wavelength, so quantum corrections
tend to increase the size of the kink.
When $ v^2 \ll y < 1$,
the quantum contribution dominates the energy,
and the kink has size $\zqu \approx 1/y$
and energy $ \Eeff \approx (y/\pi) \tscale$.

\chapter{Conclusions}

We have examined the effects of quantum corrections
on solitons in a \two--dimensional $\phi^4$ theory
with a large Yukawa coupling $y$ to fermions.
To treat the Yukawa coupling nonperturbatively,
we have solved the theory in the large-$N$ limit,
where $N$ is the number of flavors.
The solitons in this theory are kinks
which carry fermion number ranging from
$ -\half N $ to $ \half N $.
In the classical approximation,
the energy of the kink is independent of $y$,
and its size is proportional to
the scalar field Compton wavelength.
We have found that fermion loop corrections
increase the energy of the kink
and (when $y > 1$) reduce its size.
As a result of the fermion vacuum contribution,
the kink energy is bounded below by
$ (y/\pi) (N/\xcl) = Ngv/\pi $,
and its size can be as small as the fermion Compton wavelength.

When $y$ is large,
a state of $N$ fermions is expected on classical grounds
to be unstable to the formation of
a kink and anti-kink,
each carrying fermion number $ \half N$.
Quantum corrections eliminate this instability for $ v \gsim 1$
by increasing the kink/anti-kink energy.
The instability persists for $ v \lsim 1$,
but the difference in energy between the $N$ fermions
and the kink/anti-kink pair is only about 36\%
because the kink energy is proportional to the Yukawa coupling
in the large $y$ limit.

In the large-$N$ limit,
scalar loops are suppressed.
The energy of scalar field fluctuations is of order $1/\xcl$,
small compared to the classical kink energy $\sim N v^2/\xcl $.
What happens when $N$ is not large?
Will a single fermion decay into a kink/anti-kink pair
when $N=1$?
Scalar field fluctuations are still relatively unimportant
as long as $v$ is large;
$1/v^2$ is the usual semi-classical expansion parameter.
We found fermions to be stable in this regime.
Scalar corrections become more important for small $v$,
but on the other hand $1/\xcl$ is still small
relative to the quantum kink energy $y/\pi\xcl$ when $y$ is large.
It is difficult to say whether a single fermion
is unstable when $ v\lsim 1$.

Some aspects of the model discussed in this paper
are peculiar to two dimensions.
Presumably only in two dimensions can a fermion split into
a pair of solitons, each carrying fermion number ${1\over 2}$.
We expect other features of the quantum kink
to be more universal, however.
First, its energy acquires a linear dependence
on the Yukawa coupling in the strong coupling limit
through the fermion vacuum energy.
Second, for large Yukawa coupling,
fermion loop corrections tend
to reduce the size of the soliton in
the direction of the fermion Compton wavelength.
Both of these features apply not only
to the \two--dimensional solitons described in this paper,
but also to \four--dimensional (large-$N$) nontopological solitons~[\BN].

General arguments can be adduced
to suggest that these are generic features of
quantum solitons in any (large-$N$)
strongly-coupled Yukawa theory in \four~dimensions.
The fermion loop contribution to the effective action is
$ -iN \logdetD  $.
After renormalization,
its contributions to the effective potential (of order $g^4$)
and to the two-derivative term (of order $g^2$)
overwhelm the tree-level contributions when $g$ is large.\foot{
Skyrmions, in which the fermion loop contribution
to the two-derivative term vanishes after renormalization,
apparently present an exception [\MWZ,\DF,\gradient,\KPR].}
For large Yukawa coupling, therefore,
quantum solitons are determined by the fermion vacuum energy
$ (iN/T)  \logdetD $.
If this has a minimum for given boundary conditions,
the resulting configuration must have size
$ R \sim 1/gv $ (the only scale present)
and energy $ \sim (gv)^4 R^{3-d} $,
where $d$ is the dimension of the soliton.
For point-like solitons ($d=0$),
the energy is proportional to the Yukawa coupling.
Assuming that the large-$N$ restriction is only a technical one
to facilitate calculations at strong coupling,
we conjecture that these properties
hold for quantum solitons in any strongly-coupled Yukawa theory.

\ACK

I would like to thank R.~Perry for a useful conversation,
and for drawing my attention to ref.~\WK.
I am also grateful to J.~Bagger, E.~Poppitz, and S.~Mrenna
for helpful discussions.
This work has been supported by the NSF under grant PHY-90-96198.

\FIG\fig{A representative quantum kink.
The solid line shows the solution of eq.~\eequantumeom~for
the parameters $y=20$ and $v=4$.
The dashed line shows the ansatz
$ \vphi(z) = v ~\tanh(z/\zqu) $,
with $\zqu$ given by eq.~\eekinksize.
The dotted line shows the classical kink, $\zqu=1$.}

\FIG\fig{
The function $ \Deltt(t) $.
The exact fermion vacuum energy for the ansatz
$ \vphi(z) = v~\tanh(z/\zqu) $
is given by $ (N/\xcl) \, y \, \Deltt ( y \zqu ) $.}

\endpage
\refout
\figout
\end